\documentclass[prl,twocolumn,epsfig,rotate,superscriptaddress,showpacs]{revtex4}
\usepackage[dvips]{graphics}
\usepackage{graphicx}
\usepackage{amsfonts}
\usepackage{amssymb}
\usepackage{amsmath}
\usepackage{subfigure}
\newcommand{\cP}{\ensuremath{\mathcal{P}}}
\newcommand{\cT}{\ensuremath{\mathcal{T}}}
\usepackage{prettyref}
\usepackage{float}
\usepackage{amsmath}
\usepackage{amssymb}
\usepackage{esint}

\begin{document}
\title
{Intrinsic Dynamical Fluctuation Assisted Symmetry Breaking in Adiabatic Following}

\author{Qi Zhang}
\affiliation{Laboratory of Condensed Matter Theory and Computational Materials, Zhengzhou University,
Zhengzhou 450052, People's Republic of China}
\affiliation{Center for Clean Energy and Quantum Structures and School of Physics and Engineering,
Zhengzhou University, Zhengzhou 450052, People's Republic of China}
\affiliation{Centre of Quantum Technologies and Department of
Physics, National University of Singapore, 117543, Singapore}

\author{Jiangbin Gong}
\email{phygj@nus.edu.sg}
\affiliation{Department of Physics and Centre for Computational
Science and Engineering, National University of Singapore, 117542,
Singapore} \affiliation{NUS Graduate School for Integrative Sciences
and Engineering, Singapore 117597, Singapore}

\author{C.H. Oh}
\affiliation{Centre of Quantum Technologies and Department of
Physics, National University of Singapore, 117543, Singapore}
\affiliation{Institute of Advanced Studies, Nanyang Technological
University, Singapore 639798, Singapore}

\date{\today}
\begin{abstract}
 Classical adiabatic invariants in actual adiabatic processes
possess intrinsic dynamical fluctuations.
The magnitude of such intrinsic fluctuations is often thought to be negligible. This
widely believed physical picture is contested here.
For adiabatic following of
a moving stable fixed-point solution facing a pitchfork bifurcation, we show that intrinsic dynamical
fluctuations in an adiabatic process can assist in a deterministic and robust selection between
two symmetry-connected fixed-point solutions, irrespective of the rate of change of adiabatic parameters.
Using a classical model Hamiltonian also relevant to a two-mode quantum system,
we further demonstrate the formation of an adiabatic hysteresis loop in purely Hamiltonian mechanics
and the generation of a Berry phase via changing one single-valued parameter only.


\end{abstract}

\maketitle

{\it Introduction} -- Adiabatic theorem is about the dynamical behavior of a Hamiltonian system
whose parameters are changing slowly with time.  It constitutes a fundamental topic in Hamiltonian mechanics \cite{adiabatic}.
For example, Einstein was among the first
to recognize the importance of classical
adiabatic invariants in understanding quantization \cite{Gutzwiller}.
In recent years, there are still considerable interests in several aspects of
adiabatic theorem in both quantum mechanics
\cite{quantitativecondition} and classical mechanics \cite{ZhangAOP2012}.

\begin{figure}[t]
\begin{center}
\vspace*{-0.8cm}
\par
\resizebox *{7cm}{6cm}{\includegraphics*{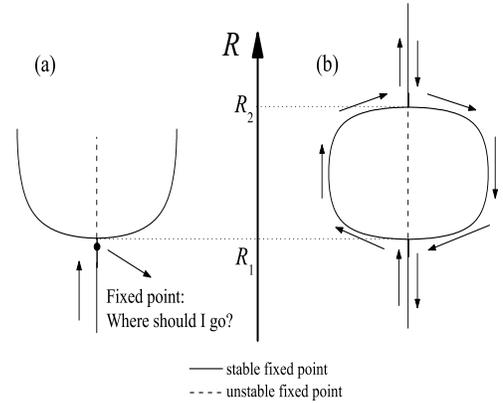}}
\end{center}
\par
\vspace*{-0.5cm} \caption{A supercritical pitchfork bifurcation
in a classical Hamiltonian system parameterized by $R$. The solid (dashed)
lines represent the phase space
locations of stable (unstable) fixed points as a function of $R$. (a) When $R$
exceeds $R_1$, a stable fixed point splits into two stable and one unstable fixed points. One then wonders
which branch the system will land on as $R$ adiabatically increases beyond $R_1$.
(b) The bifurcated fixed points merge back when $R$ exceeds $R_2$.
If $R$ first increases from $R<R_1$ to $R>R_2$ and later returns to $R<R_1$, then an adiabatic
hysteresis loop may be formed. }\label{fig1}
\end{figure}

Classical adiabatic following is the subject of this study, but our findings are also relevant to certain quantum systems.
We start from the fact that classical adiabatic theorem is not an exact theorem: an actual adiabatically evolving trajectory
fluctuates around an idealized solution predicted by the adiabatic theorem.  Adiabatic invariants hence possess
intrinsic dynamical fluctuations (IDF's) \cite{ZhangAOP2012,Berry1996, adam, mag,liu}.
The magnitude of such fluctuations, typically proportional to the rate of change of adiabatic parameters,
becomes extremely small in truly slow adiabatic processes.  So except for special quantities that can accumulate IDF during an adiabatic process \cite{liu, ZhangAOP2012} (e.g., in calculations of dynamical angles), IDF does not seem to be interesting or physically relevant.  As shown in this Letter via both theory and computational examples, this perception is about to change.

Bifurcation phenomenon is ubiquitous in nonlinear systems and it is of fundamental interest to many topics in physics, among which we mention
localization-delocalization phase transitions and symmetry breaking \cite{wubiao,wubiao2,luis,leech,ZhangPRA2008,liu2}.
Here we consider the adiabatic following of a stable fixed-point solution, which,
as a result of a varying adiabatic parameter, moves towards
a supercritical pitchfork bifurcation.  As schematically shown in Fig. 1(a), two new stable fixed points and one unstable fixed point
emerge after the adiabatic parameter (denoted $R$)
slowly passes the bifurcation point at $R=R_1$.  It is then curious to know among the three fixed points,
which fixed-point solution the system will land on and whether the selection is predictable.
It is found, both theoretically and numerically,
that IDF is crucial for a {\it deterministic} and {\it robust} selection between the symmetry-connected pair of stable fixed-point solutions, regardless of how slow the adiabatic
process is.  As such, through crossing the bifurcation point, a tiny IDF
is amplified to a macroscopic level after the bifurcation: it determines the fate of the trajectory afterwards by ``forcing" the system to
make a selection between symmetry-breaking solutions.  We term this as deterministic symmetry breaking,  because there is no need for external
noise to initiate the symmetry-breaking.  The selection process is robust because, unlike symmetry-breaking selection processes
studied previously \cite{liu2}, it is independent of the dynamical details.
  Figure 1(b) depicts an interesting situation that involves a second bifurcation at $R=R_2$, after which the three fixed-point solutions merge back to one stable fixed point.  We shall show that this case may allows us to generate a Berry phase
  by manipulating one single-valued parameter only.

{\it Preliminaries} -- We recently developed a general description of IDF in classically integrable systems \cite{ZhangAOP2012}, which can be reduced to a rather simple form
for  stable fixed-point solutions in phase space.
In particular, let us consider a one-dimensional Hamiltonian $H(q,p;R)$, with $(q,p)$ being the canonical variables, $R$ a system parameter to be tuned slowly, and
its stable fixed point solutions denoted by $[\bar{q}(R),\bar{p}(R)]$.  According to the traditional picture offered by classical adiabatic theorem, a stable phase space fixed point has zero action, so when $R$ varies slowly, the system must retain its zero action as an adiabatic invariant and therefore must follow the instantaneous fixed point
$[\bar{q}(R),\bar{p}(R)]$. This is however a picture without IDF.
The actual time evolving state $(q,p)$ deviates from  $[\bar{q}(R),\bar{p}(R)]$,
\begin{equation}
q=\bar{q}+\delta q,\ \  p=\bar{p}+\delta p.
\end{equation}
where $(\delta q, \delta p)$ are IDF on top of the idealized adiabatic solution $[\bar{q}(R),\bar{p}(R)]$ with $R=R(t)$.
It is straightforward to see why $(\delta q, \delta p)$ has to be nonzero: were it indeed zero,
then by definition of a fixed point, we have $\frac{\partial H(q,p,R)}{\partial q}= \frac{\partial H(q,p,R)}{\partial p}=0$,
indicating that the current state $(q,p)$ cannot evolve and hence the system can never do adiabatic following with a
moving fixed-point solution.
This indicates that nonzero $(\delta q, \delta p)$ is not due to nonadiabaticity. Rather, it is intrinsic and
must exist for adiabatic following to occur.
Our theory applied to this case~\cite{ZhangAOP2012} gives (see also \cite{liu})
\begin{equation} \label{fluctuation}
\left(\begin{array}{c}\langle\delta{p}\rangle\\
\langle\delta {q}\rangle\end{array}\right)= {\Gamma}^{-1}
 \left(\begin{array}{c}\frac{\partial \bar{p}}{\partial R}\\
 \frac{\partial \bar{q}}{\partial R}\end{array}\right) \cdot
 \frac{dR}{dt},
\end{equation}
where
\begin{equation} \label{Gamma}
\Gamma=\left(\begin{array}{cc}-\frac{\partial^2H}{\partial q\partial
p}&-\frac{\partial^2 H}{\partial q\partial q}
\\\frac{\partial^2H}{\partial p
\partial p}&\frac{\partial^2H}{\partial
p \partial q}
\end{array}\right)_{p=\bar{p},q=\bar{q}},
\end{equation}
and $\langle \cdot\rangle$ denotes an average over all possible initial conditions of $(\delta q, \delta p)$.
As seen from Eq.~(\ref{fluctuation}), so long as $\Gamma^{-1}$ exists, then the scale of IDF is proportional to $\frac{dR}{dt}\equiv V$, the speed of adiabatic manipulation.  The magnitude of $(\delta q,\delta p)$ is then deceptively small (for a sufficiently small $V$), but nonzero in general.

\begin{figure}[t]
\begin{center}
\vspace*{-1.5cm}
\par
\resizebox *{8.4cm}{6.5cm}{\includegraphics*{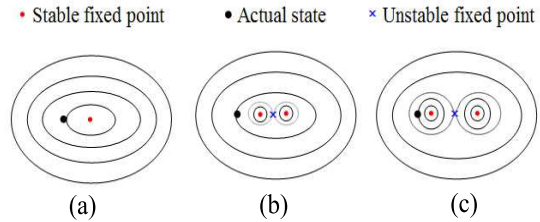}}
\end{center}
\par
\vspace*{-1.2cm} \caption{(color online) Impact of
intrinsic dynamical fluctuations on the crossing of a pitchfork bifurcation.
(a) The system's actual state is on the left of
the instantaneous fixed point. (b) Immediately after the bifurcation, the
actual state is still shifted to the left of all the fixed points.
(c) As two stable fixed points split further, the actual state gets trapped by, and starts
to adiabatically follow, the stable fixed-point solution on the left.
}\label{fig2}
\end{figure}

We now examine how nonzero $(\delta q,\delta p)$ impacts the adiabatic following of $[\bar{q}(R),\bar{p}(R)]$ that eventually undergoes
a supercritical pitchfork bifurcation at $R=R_1$.
First, right before the adiabatic parameter $R$ reaches the bifurcation point $R_1$, the system's
actual state $(q,p)$ must deviate from the instantaneous solutions $[\bar{q}(R),\bar{p}(R)]$
by $(\delta q, \delta p)$.  Without loss of generality and based on Eq.~(\ref{fluctuation}), the actual time-evolving state $(q,p)$
is assumed to be slightly shifted to the left side of the instantaneous fixed point $[\bar{q}(R),\bar{p}(R)]$.  This is illustrated in Fig.~2(a), where
the periodic orbits (associated with a fixed $R$) around $[\bar{q}(R),\bar{p}(R)]$ are also shown.
The second stage is illustrated by Fig.~2(b), where $R$ only slightly exceeds $R_1$. There the
bifurcation already occurs but the actual state does not ``feel" the bifurcation yet: it continues to stay on the left side
of all the three new fixed points.
So the actual left-shifted state is located on an orbit (if $R$ were fixed) surrounding all the fixed points (this is confirmed in our numerical studies).
The line shown in Fig.~2(b) passing through the
unstable fixed point represents the separatrix.  In the last stage, $R$ further increases, the two stable fixed points split further, with
the stable fixed point moving to the left capturing the actual state [see Fig. 2(c)].  During  the ensuing adiabatic following,
the actual state then adiabatically follows the instantaneous fixed point on the left.
Clearly then, if, as $R$ approaches a bifurcation point, IDF can induce a definite
shift (to the left or the right) of the actual state with respect to  $[\bar{q}(R),\bar{p}(R)]$,
then the system will be trapped,  deterministically, by one of
the two stable fixed-point solutions after the bifurcation.
Note that, exactly because of IDF, the actual state always stays away from the vicinity of the unstable fixed point (the extremely
slow part of a separatrix). As a result the diverging time scale associated with
the whole separatrix does not affect adiabatic following here.
This understanding will be confirmed in our following numerical experiments.

{\it Model Hamiltonian} -- Here we turn to a concrete Hamiltonian system with supercritical pitchfork bifurcations.  Specifically,
we choose
\begin{equation} \label{HamiltonianC}
H=-\frac{c}{2}q^2-R\sqrt{1-q^2}\cos(p)+\Delta\sqrt{1-q^2}\sin(p)
\end{equation}
in dimensionless units,
with $c>\Delta>0$.
Interestingly, this system has two bifurcation points at $R_1=-\sqrt{c^2-{\Delta^2}}$ and $R_2=\sqrt{c^2-\Delta^2}$.
Define $\eta=\sqrt{1-(R^2+\Delta^2)/c^2}$ and $\mu=\arctan(-\Delta/R)$. In the negative regime of $R$,
the stable fixed point is $(\bar{q},\bar{p})=(0,\mu-\pi)$ for $R<R_1$, which then bifurcates into two stable
fixed points at $(\mp \eta, \mu-\pi)$.
In the positive regime of $R$, the two stable fixed points are at $(\mp \eta, \mu)$ for $R<R_2$ and then
merge back to one stable fixed point at $(0, \mu)$.  Note that this model Hamiltonian is invariant under the joint operation of
space reflection (\cP) $q\rightarrow -q$ and time reversal (\cT) $t\rightarrow -t$.  Therefore, for cases with only one stable fixed point, the solution itself has the same \cP\cT\ symmetry, and for cases with two stable fixed points, the pair are transformed to each other under the \cP\cT\
operations, with each individual fixed point being a symmetry-breaking solution.
The phase space locations of these stable fixed-point solutions for various fixed
values of $R$ are shown by the solid lines in Fig.~3.

Another motivation to choose $H$ in Eq.~(\ref{HamiltonianC}) is that it describes a two-mode many-body
quantum system on the mean-field level.  Consider the following mean-field Hamiltonian in dimensionless units ($\hbar=1$),
\begin{equation} \label{Hamiltonian}
\hat{H}_{\text{m}}=\left(\begin{array}{cc}c(|b|^2-|a|^2)&-R-i\Delta
\\-R+i\Delta&-c(|b|^2-|a|^2)
\end{array}\right),
\end{equation}
where $a$ and $b$ are quantum amplitudes on two modes, $c$ represents the self-interaction strength,
and $R \pm i \Delta$ represents inter-mode coupling.
By rewriting $a=|a| e^{i\phi_a}$, $b=|b| e^{i\phi_b}$, $p=\phi_b-\phi_a$, and $q=|b|^2-|a|^2$, the time
evolution of $q$ and $p$, as obtained from the Schr\"{o}dinger equation for this quantum model, becomes precisely
that under the classical model Hamiltonian $H$ in Eq.~(\ref{HamiltonianC}). The fixed points
of $H$ become eigenstates of $\hat{H}_{\text{m}}$;  the adiabatic following when crossing a bifurcation
for $H$ is mapped to the issue of adiabatic following as degenerate eigenstates of
$\hat{H}_{\text{m}}$ emerge.
One possible realization of $\hat{H}_{\text{m}}$ is a Bose-Einstein condensate (BEC) in a double-well potential, with
the imaginary coupling constant $\Delta$ implemented via phase imprinting on
one well \cite{Denschlag}.  $\hat{H}_{\text{m}}$ may also be realized in nonlinear optics by using
two nonlinear optical waveguides with biharmonic longitudinal modulation of the refractive
index ~\cite{Kartashov}.   So our detailed results below are relevant to both classical and quantum physics.

\begin{figure}[t]
\begin{center}
\vspace*{-0.cm}
\par
\resizebox *{9cm}{6.5cm}{\includegraphics*{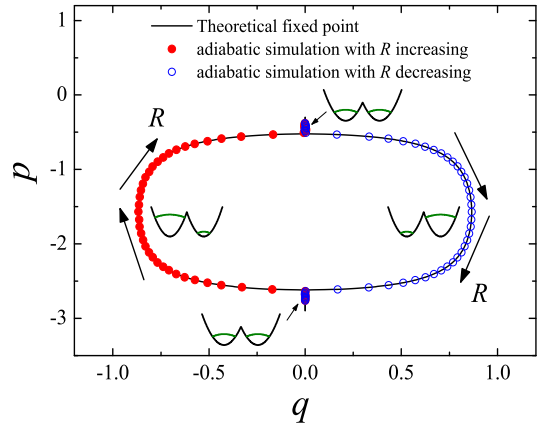}}
\end{center}
\par
\vspace*{-0.5cm} \caption{(color online) Solid lines depict phase space locations of fixed-point solutions for our model Hamiltonian in Eq.~(\ref{HamiltonianC}) as values of $R$ are scanned. The solid and empty dots represent the actual states of the system in  numerical experiments of an adiabatic process, where
 $R$ is first increased from $R=-0.25<R_1$ to $R=0.25>R_2$, and then returns to $R=-0.25$, at a constant speed $V=10^{-6}$. Long arrows indicate the moving direction of the actual states as $R$ changes. The four double-well plots schematically show the symmetry-breaking consequence in the context of a two-mode BEC Hamiltonian. $c=0.2$, $\Delta=0.1$. Variables plotted here and in Fig.~4 are in dimensionless units.
 }\label{fig3}
\end{figure}

{\it Theory and numerical experiments} -- Applying the theory of IDF to the Hamiltonian in Eq.~(\ref{HamiltonianC}), one obtains
$\langle \delta p\rangle  =0$ and
$\langle \delta q\rangle = -\frac{\Delta}{\sqrt{R^2+\Delta^2}}\frac{1}{R^2+\Delta^2-c\sqrt{R^2+\Delta^2}}\frac{dR}{dt}$ for $R<R_1$~\cite{notedivergence}.
Therefore, as $R$ increases from $R<R_1$, $\frac{dR}{dt}=V>0$, and  $\langle \delta q\rangle$ is definitely negative.
Returning to the phase space plot in Fig.~3, this means that
if $R$ increases from $R<R_1$, the actual state (on average, to be more precise) is slightly left-shifted from the fixed point at $(0,\mu-\pi)$.
So after passing the bifurcation at $R=R_1$, the system is expected to adiabatically follow the left symmetry-breaking
solution $(\bar{q},\bar{p})= (-\eta, \mu-\pi)$ for $R_1<R<0$ and $(\bar{q},\bar{p})= (-\eta, \mu)$ for $0<R<R_2$.
Similar theoretical results show that, if $R$ decreases from $R>R_2$, then
the actual state must be slightly right-shifted from the fixed point at $(0,\mu)$, a prediction consistent with the above-mentioned \cP\cT\ symmetry of the system.
So after
crossing the bifurcation point at $R=R_2$, the system should adiabatically follow the other symmetry-breaking
solution $(\bar{q},\bar{p})= (\eta, \mu)$ for $0<R<R_2$ and $(\bar{q},\bar{p})= (\eta, \mu-\pi)$ for $R_1<R<0$.
As shown in Fig.~3 (solid or empty dots), our numerical results based only on Hamilton's equation of motion
confirm this prediction. For the results shown
we have set $V=10^{-6}$ to ensure a slow process. Indeed, at all times the difference between the actual states (dots) and the instantaneous fixed points (solid lines) is invisible to our naked eyes. Yet, the small IDF does assist in a symmetry-breaking choice regarding which of two stable fixed-point solutions is adiabatically followed by the system. In the language of $\hat{H}_{\text{m}}$, after the system passes the bifurcation at $R=R_1$ owing to an increasing $R$, $q=|b|^2-|a|^2$ becomes appreciably negative, so one mode develops
more population than the other, signaling a clear delocalization-localization transition induced by IDF.
The opposite population imbalance occurs when the system passes the bifurcation at $R=R_2$ with a decreasing $R$.
Furthermore, joining these two manipulation steps together so that $R$ returns to its very original value in the end (which is already the case in Fig.~3), we clearly observe the formation of an adiabatic ``hysteresis" loop in phase space.  That is, by increasing and then decreasing $R$, a navigation loop in phase space is formed because the adiabatic following in the forward step and backward step lands on different symmetry-breaking branches \cite{note}.

There is one subtle point to be clarified: Our theory of IDF [see (Eq.~\ref{fluctuation})] is about quantities $\langle\delta q\rangle$ averaged over all initial conditions of $(\delta q,\delta p)$, but what determines the adiabatic following is the actual $\delta q$ in a single process.
The Supplementary Material contains a detailed analysis on this point. In particular,
we show that if initially there is a deviation of  $\delta q$ from $\langle \delta q\rangle$,  then the difference $\delta q- \langle \delta q\rangle$ oscillates with time and later, as $R$ approaches the bifurcation point, this deviation becomes negligible as compared with $\langle\delta q\rangle$.  It is for this reason
that the definite sign of $\langle\delta q\rangle$ is equivalent to the definite sign of $\delta q$, which hence justifies our theory based on $\langle\delta q\rangle$.

{\it Berry phase generation via one single-valued parameter} -- As a final interesting concept, we discuss the generation of a Berry phase
using one single-valued adiabatic parameter $R$.  This is made possible by IDF and bifurcations.
In particular, the navigation loop in Fig.~3 shows that the time-evolving states of $\hat{H}_{\text{m}}$ trace out a nontrivial geometry
after increasing $R$ from $R<R_1$ to $R>R_2$ and then returning $R$ to its initial value.  Let $|\psi_{\text{left}}\rangle$ and $|\psi_{\text{right}}\rangle$ be the eigenstates of $\hat{H}_{\text{m}}$ (with $R_1<R<R_2$) mapped from the left and right fixed-point solutions of $H$.
Assuming exact adiabatic following with the instantaneous adiabatic eigenstates, the Berry phase generated along the navigation loop
is analytically found to be
\begin{eqnarray} \label{Berry}
\beta_{\text{Berry}}
&=&i \int_{R_\text{1}}^{R_\text{2}}
dR\left(\langle\psi_{\text{left}}|\frac{d}{dR}
|\psi_{\text{left}}\rangle-\langle\psi_{\text{right}}|
\frac{d}{dR}|\psi_{\text{right}}\rangle\right) \nonumber \\
&=&\pi(1-\Delta/c).
\end{eqnarray}
In our numerical experiments, we choose to integrate the Berry connection using the actual states during the physical process
(this simple method will not account for the nonlinear Berry phase correction studied in Ref.~\cite{liu}). It is found that for
the shown regime in Fig.~4, the agreement between theory and simulations is excellent, with tiny but visible differences.  Such visible
differences remind us that a direct numerical integration of the Berry connection along the actual time evolution path
is not necessarily reliable because it could accumulate the effect of IDF \cite{ZhangAOP2012}.
Nevertheless, the fair agreement shown in Fig.~4 confirms our analytical result, demonstrates the feasibility of Berry phase generation using only one
single-valued adiabatic parameter, and verifies from another angle that IDF is important for understanding adiabatic following in the presence of bifurcation.

\begin{figure}[t]
\begin{center}
\vspace*{-1.5cm}
\par
\resizebox *{6.9cm}{5.65cm}{\includegraphics*{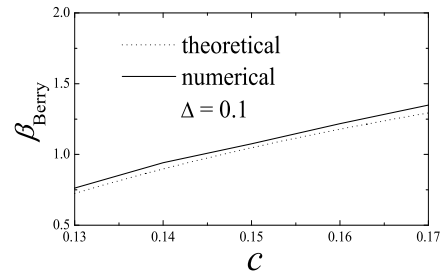}}
\end{center}
\par
\vspace*{-0.5cm} \caption{Analytical and numerical Berry phase results vs $c$, for the quantum system
described in Eq.~(\ref{Hamiltonian}) with $\Delta=0.1$.  The  theoretical
results are from Eq.~(\ref{Berry}), while the numerical results
are obtained directly by integrating the Berry connection along the actual path
of adiabatic evolution. }
\end{figure}

{\it Conclusion} --  Bifurcation greatly amplifies subtle intrinsic fluctuations that are beyond classical adiabatic invariants.
This leads to a selection rule regarding which of two symmetry-connected stable fixed-point solutions may be adiabatically followed.
In cases of multiple bifurcation points,  adiabatic hysteresis loops in phase space and the generation of Berry phase by manipulating one single-valued parameter are also shown to be possible.
Our findings are of fundamental interest to both classical systems
and quantum many-body systems on the mean-field level.  The implications of this work for symmetry breaking in fully quantum
many-body systems should be a fascinating topic in our future studies.

\vspace{0.5cm}
{\begin{center} {\bf Appendix} \end{center}}

In this Appendix we discuss the difference between $\langle \delta q\rangle$ and $\delta q$, using the same notation as in the main text.
It is important to clarify this point because our theory of intrinsic dynamical fluctuations (IDF) is about
the quantity $\langle\delta q\rangle$ averaged over all initial conditions of $(\delta q, \delta p)$, but, as indicated in our main text, what determines the symmetry-breaking adiabatic following is the actual $\delta q$ in a single process without
the averaging.  Specifically, we need to show that, before the system crosses the bifurcation point,
the (positive or negative) sign of $\delta q$ in an actual process is always the same as the sign of $\langle\delta q\rangle$ predicted by our theory. Without loss of generality we choose to discuss the case $R<R_1$ as an example.

For our model Hamiltonian, the $\Gamma$ matrix evaluated at the fixed point is found to have zero diagonal elements.  Hence we can write it in the following form
\begin{equation}
\Gamma=\left(\begin{array}{cc}0&-A
\\B&0
\end{array}\right).
\end{equation}
Next we expand the Hamilton's equation of motion to the first order of $\delta q$ and $\delta p$, we have
\begin{eqnarray}
\left(\begin{array}{c}\frac{dp}{dt}\\\frac{dq}{dt}\end{array}\right)&=&\left(\begin{array}{cc}0&-A
\\B&0
\end{array}\right)\left(\begin{array}{c}\delta p\\ \delta q\end{array}\right)\nonumber \\
&=& \left(\begin{array}{cc}0&-A
\\B&0
\end{array}\right) \left(\begin{array}{c}p-\bar{p}\\ q-\bar{q}\end{array}\right),
\end{eqnarray}
where the instantaneous fixed point $(\bar{q},\bar{p})$ is located at $[0, \arctan(-\Delta/R)-\pi]$. Note that
in general the off-diagonal elements $A$ and $B$ are $R$-dependent. Nevertheless, to gain physical insights and
to develop a simple analytical result from the above equation, we consider a small time segment
during which the $\Gamma$ matrix can be regarded as a constant, and $\bar{p}$ moves at a rate $M$ due to an increasing $R$.
We then obtain
\begin{eqnarray} \nonumber
\frac{d p}{dt}&=&-Aq; \\
\frac{d q}{dt}&=&Bp-BMt,
\end{eqnarray}
a first-order differential equation that can be integrated directly. Upon direct integration
we find the analytic solution
\begin{eqnarray} \label{simplesolution} \nonumber
q&=&D\cos(\sqrt{AB}t)-\frac{M}{A} \\
p&=&-AD\sin(\sqrt{AB}t)+Mt,
\end{eqnarray}
where $D$ is one integration constant.  This then indicates
\begin{eqnarray}
\delta q&=&D\cos(\sqrt{AB}t)-\frac{M}{A}; \nonumber \\
\delta p&=&-AD\sin(\sqrt{AB}t). \label{final}
\end{eqnarray}
Equation (\ref{final}) clearly shows that $\delta q$ and $\delta p$ are oscillating solutions. In particular,
the oscillation amplitude in $\delta q$ is seen to be the initial difference between $\delta q$ and the averaged quantity
$\langle\delta q\rangle=-M/A$ (for the concerned time segment). Even more importantly, as the system approaches the bifurcation point at $R=R_1$, the term $-M/A$
is the dominating term because $|A|$ sharply decreases for $R$ approaching $R_1$.  This makes it clear that the sign of $\delta q$ must agree
perfectly with the sign of $\langle\delta q\rangle$.

\begin{figure}[t]
\begin{center}

\vspace*{-0.cm}
\par
\resizebox *{7.5cm}{6.5cm}{\includegraphics*{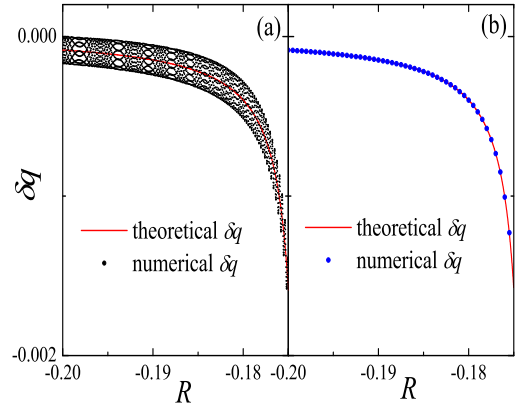}}
\end{center}
\par
\vspace*{-0.5cm} \caption{(color online) Comparison between numerical results of $\delta q$ and theoretical $\langle \delta q\rangle$ obtained from
 theory of IDF. System parameters are $c=0.2$, $\Delta=0.1$, and $\frac{dR}{dt}=10^{-6}$. (a) Initially the canonical variables $(q,p)$ are
 set exactly on the instantaneous fixed point. The initial $\delta q$ is hence zero, which differs from the theoretical nonzero $\langle \delta q\rangle$.
 (b) Initially $\delta p$ and $\delta q$ are precisely chosen according to the theory of IDF (see the main text). In this case, $\delta q$ is seen to
 agree with $\langle \delta q\rangle$ at all times before the bifurcation, which is about to occur at $R=-0.173$ for parameters used here.}
 \label{figsupp}
\end{figure}

To verify this theoretical understanding, we performed numerical experiments and quantitatively compare in Fig.~\ref{figsupp}
the numerically found $\delta q$ with $\langle\delta q\rangle$ obtained from the theory of IDF (see the main text), for an increasing $R$ ($R<R_1$).
As seen in Fig.~\ref{figsupp}(a), the actual numerical $\delta q$ indeed oscillates around theoretical $\langle\delta q\rangle $. As $R$ approaches closer to $R_1=-0.173$ (beyond the shown regime of $R$ in Fig.~\ref{figsupp}),
the oscillations of $\delta q$ around $\langle\delta q\rangle $ remain small and yet $\langle\delta q\rangle $ becomes more and more negative.
Therefore, the sign of $\delta q$ is always identical with the sign of  $\langle\delta q\rangle $ when the system is close to bifurcation crossing.
As a second confirmation of our insights above, Fig.~\ref{figsupp}(b) shows that, if initially we set $\delta q=\langle\delta q\rangle $ and $\delta p=\langle\delta p \rangle$, then the ensuing time dependence of $\delta q$ as found from our numerical calculation agrees exactly with that from the theory of IDF.

We finally note that, the ultimate reason why our analytical solution above for $(\delta q, \delta p)$ during a small time segment is already so useful
is again linked with adiabatic following. That is, since $R$ changes slowly,
the off-diagonal elements of $\Gamma$ and $M$ will all change slowly, and hence the equation of motion for $\delta q$, or for the deviation
$\delta q-\langle \delta q\rangle $, is expected to adiabatically follow the solution given in Eq.~(\ref{final}).   This is interesting because it suggests a theory
of higher-order fluctuations in the intrinsic fluctuations.


\begin{thebibliography}{99}
\bibitem{adiabatic} P.A.M. Dirac, Proc. R. Soc. {\bf 107}, 725 (1925); M. Born and V. A. Fock, Zeitschrift f\"ur Physik A {\bf 51}, 165 (1928).
\bibitem{Gutzwiller}M.~C. Gutzwiller, {\it Chaos in Classical and Quantum Mechanics} (Springer-Verlag, New York 1990), pp208-211.
\bibitem{quantitativecondition} K. P. Marzlin and B. C. Sanders, Phys. Rev. Lett. {\bf 93},
160408 (2004); Z.~Y.~Wu and H.~Yang, Phys. Rev. A {\bf 72}, 012114 (2005); R. Mackenzie, E. Marcotte, and H. Paquette,
Phys. Rev. A {\bf 73}, 042104 (2006); J. Ma, Y. P. Zhang, E. G. Wang, and B. Wu, Phys. Rev.
Lett. {\bf 97}, 128902 (2006); D. M. Tong, K. Singh, L. C. Kwek, and C. H. Oh, Phys. Rev.
Lett. {\bf 95}, 110407 (2005); D. M. Tong, K. Singh L. C. Kwek, and C. H. Oh, Phys. Rev. Lett. {\bf 98}, 150402 (2007); D. M. Tong, Phys. Rev. Lett. {\bf 104}, 120401 (2010).
\bibitem{ZhangAOP2012} Q.~Zhang, J.~B.~Gong, and C.~H.~ Oh, Ann. Phys. {\bf 327}, 1202 (2012); Q.~Zhang, J. Phys. A {\bf 45}, 295302 (2012).
\bibitem{Berry1996} M.~V.~Berry and M.~A.~Morgan, Nonlinearity {\bf 9}, 787 (1996).
\bibitem{adam}A.~D.~A.~M.~Spallicci, A. Morbidelli, and G. Metris, Nonlinearity {\bf 18}, 45 (2005).
\bibitem{mag}M.~V.~Berry and J.~M.~Robbins, Proc. Roy. Soc. Lond. A{\bf 442}, 641 (1993). 
\bibitem{liu} J.~Liu and L.B. Fu, Phys. Rev. A {\bf 81}, 052112 (2010). 
\bibitem{wubiao} B.~Wu and Q.~Niu, \pra{\bf 61}, 023402 (2000).
\bibitem{wubiao2} X. Luo, Q. Xie, and B. Wu, Phys. Rev. A {\bf 77}, 053601 (2008).
\bibitem{luis}L.~Morales-Molina and S.~Flach, New J. Phys. {\bf 10}, 013008 (2008).
\bibitem{leech} C.~H.~Lee, \prl{\bf 102}, 070401 (2009); C.~H.~Lee, W.~H.~Hai, L.~Shi, and K.~L.~Gao,  \pra{69}, 033611 (2004).
\bibitem{ZhangPRA2008} Q.~Zhang, P.~H\"anggi, and J.~B.~ Gong, Phys. Rev. A {\bf 77}, 053607 (2008).
\bibitem{liu2} D.~F.~ Ye, L.~B.~Fu, and J.~Liu, \pra{\bf 77}, 013402 (2008); L.~B.~Fu, D.~F~.Ye, C.~H.~ Lee, W.~P. Zhang, and J.~Liu, \pra{\bf 80}, 013619 (2009).
\bibitem{Denschlag} J. Denschlag {\it et al},  Science {\bf 287}, 97 (2000).
\bibitem{Kartashov} Y.~V.~Kartashov and V.~A.~Vysloukh, Opt. Lett. {\bf 34}, 3544 (2009).
\bibitem{notedivergence} The expression
of $\langle \delta q\rangle$ diverges at $R=R_1$, because in our model, $\Gamma^{-1}$ diverges at isolated points $R_1$ or $R_2$.
 Such unphysical divergence can be removed
 by considering a second order expansion of $\delta q$ and $\delta p$ (which then predicts that $\delta q$ and $\delta p$
 can be of the order of $\sqrt{|dR/dt|}$). This procedure is unnecessary here because we do not need to estimate the precise magnitude
 of $\langle \delta q\rangle$ to make symmetry-breaking predictions (rather, only its sign is crucial).
\bibitem{note}  A preliminary observation as a side result in a driven two-mode system in Ref.~\cite{ZhangPRA2008} (coauthored by two of us)
also suggested the possibility of adiabatic hysteresis loops. However, therein the origin of symmetry breaking
was not well understood due to the lack of connection with IDF.










\end{thebibliography}
\bibliographystyle{apsrev}

\end{document}